Compositional Evolution of the Anti-phase Stripe Superstructure in $Bi_2Sr_{2-x}La_xCuO_6$ ($0<x\leq1.1$) Revealed by Transmission Electron Microscopy

Z. Chen*, Y. Y. Peng*, Z. Wang, Y. J. Song, J. Q. Meng, X. J. Zhou, J. Q. Li[†]

Beijing National Laboratory for Condensed Matter Physics, Institute of Physics, Chinese Academy of Sciences, Beijing 100190, People's Republic of China

Abstract

The detailed structural properties of La-doped $Bi_2Sr_{2-x}La_xCuO_6$ (La-Bi2201, $0\leq x\leq1.1$) have been studied by transmission electron microscopy (TEM). The well-known incommensurate supermodulation $\mathbf{q}_1$ in the Bi-based cuprates evolves from a monoclinic superstructure in the pristine Bi2201 phase to an orthogonal one in La-Bi2201($x=0.73$) phase. The *b*-component of the modulation vector ($\mathbf{q}_1$) for $x=0.25$ sample is about $0.24\mathbf{b}^*$ and increases slightly to $0.246\mathbf{b}^*$ in $x=0.84$ sample, while it increases significantly to $0.286\mathbf{b}^*$ for $x=1.10$ sample. We have revealed a new anti-phase stripe superstructure in all the La-doped Bi2201 samples, giving rise to a new modulation with a vector $\mathbf{q}_2$. This $\mathbf{q}_2$ modulation is directly evolved from the orthogonal modulation $\mathbf{q}_1$ through an addition of the anti-phase structure so that its vector $q_2$ equals to $q_{1b}/2$. We also discussed the implication of these structural studies on the electronic structure by angle-resolved photoemission spectroscopy (ARPES) experiments in La-Bi2201.

*These authors contribute equally to this work.
[†]Correspondence should be addressed to J. Q. Li (ljq@aphy.iphy.ac.cn)

(Some figures may appear in color only in the online journal)

1. Introduction

The $Bi_2Sr_2CuO_6$ (Bi2201) superconductor is an ideal system to study the physical properties, electronic structure and mechanism of high-temperature superconductivity [1-7]. First, it has a simple crystal structure that contains only single $CuO_2$ layer in a unit cell which avoids possible complications from two-layered or multi-layered compounds which may give rise to multiple Fermi surface sheets. Second, by replacing $Sr^{2+}$ with some rare-earth ions ($R^{3+}$) combined with annealing process, one can obtain a wide range of doping levels that spans from heavily underdoped, underdoped, optimally-doped to overdoped region. This is beneficial for investigating systematic evolution of the electronic structure and physical properties, thus providing clues for understanding the physics and mechanism of superconductivity. Third, it has relatively low superconducting transition temperature ($T_c$). This facilitates the study of normal state properties such as the nature of the pseudogap state, while reducing the effect of thermal broadening on the electronic spectra. Fourth, it can be easily cleaved to get a clean and smooth surface, which is necessary for experiments like angle-resolved photoemission spectroscopy (ARPES) and scanning tunneling microscopy/spectroscopy (STM/STS).

Many important insights have been obtained from studying Bi2201 system. In the La and Pb co-doped $Bi_{2-y}Pb_ySr_{2-z}La_zCuO_{6+x}$ samples, STM found a static, no-dispersive and strong doping dependence 'checkerboard'-like electronic modulation in a broad doping regime which was suggested to originate from charge-density-wave formation [7]. Novel Fermi surface topologies, such as Fermi arcs in pseudogap state, signatures of particle-hole symmetry breaking in the pseudogap state [8-10] and hints of a "pseudogap phase transition" in Bi2201 [4] have been observed by ARPES. Recently, Fermi pocket has been revealed in underdoped $Bi_2Sr_{2-x}La_xCuO_6$ (La-Bi2201), which coexists with the Fermi arcs [1]. It is under debate whether such a Fermi pocket is intrinsic or is produced from another modulation $q_2$ in Bi2201 in addition to the regular incommensurate modulation $q_1$. A systematic investigation of the structural properties in Bi2201 and its variants is a pre-requisite to understand the electronic structure revealed by ARPES, STM/STS and other techniques.

The structure of the pristine Bi2201 has been studied extensively both by x-ray diffraction (XRD) and transmission electron microscopy (TEM) [11-15]. As shown in figure 1, the basic structure of Bi2201 is an orthorhombic layered structure with cell parameters of a=5.362Å, b=5.374Å, c=24.622Å [11]. An incommensurate superstructure in the *bc* plane has been identified by many high-resolution TEM (HRTEM) and electron diffraction studies (in [13], the modulation vector is $q_1 \approx 0.217b^* + 0.62c^*$). The origin of this supermodulation is mainly attributed to the distortions of the BiO bilayers [12, 16, 17]. Substitutions in Bi or Sr sites usually modify the distortions of the BiO bilayers, and may lead to a change of the modulation vector $q_1$. There have been several TEM studies on La-Bi2201 with different compositions. In [12], the electron diffraction patterns (DPs) of $Bi_2Sr_{2-x}La_xCuO_6$ showed an increase of the *b*-component of the modulation vector, from $0.213b^*$ for x=0 to $0.294b^*$ for x=1. However, they only gave DPs along the [100] zone axis but did not provide DPs along the [001] zone axis which are more closely related to the $CuO_2$ planes. There are other TEM works that either covered only a limited doping concentrations [18] or on samples with a large deviation from the ratio Bi/Sr=1 [19]. So far, a systematic TEM study on the microstructure evolution of La-Bi2201 over a whole range of x ($0<x\leq1.1$) is not available. Such a study is highly desirable to understand the electronic structure and physical properties of La-Bi2201 that have been probed by many other techniques [1, 7-10, 20, 21].

Recent low-energy electron diffraction (LEED) study, combined with ARPES experiments, found a new superstructure $\mathbf{q}_2$ (called LEED $\mathbf{q}_2$ hereafter) in the underdoped La-Bi2201 samples, in addition to the usual superstructure modulation $\mathbf{q}_1$ [20]. Since LEED is a surface sensitive technique, this LEED $\mathbf{q}_2$ should be related to a modulation on the surface of La-Bi2201. This LEED $\mathbf{q}_2$ was proposed to be structural [20] and was attributed to be the origin of the Fermi pocket observed in ARPES [1]. A further study combining surface-sensitive probes like ARPES and LEED, and bulk-sensitive probes like resonant (REXS) and non-resonant (XRD) x-ray diffraction, suggested that the new incommensurate superstructure $\mathbf{q}_2$ behaves differently on the surface and in the bulk [21]. This leaves many issues to be addressed. What are the main characters of the new superstructure $\mathbf{q}_2$, such as its $\mathbf{q}$-vector? How does it evolve with doping and temperature? What is the exact nature of this new $\mathbf{q}_2$—whether it is a bulk character or surface one? In order to clarify these problems, we carried out a systematic TEM studies for a series of La-Bi2201 samples covering a wide composition range of x ($0<x\leq1.1$).

2. Experimental details

Poly-crystalline parent phase of $Bi_2Sr_2CuO_6$ (Bi2201) was synthesized by the conventional solid-state reaction method. All the La-doped $Bi_2Sr_{2-x}La_xCuO_6$ (La-Bi2201) single crystalline samples used in our TEM studies were grown through the traveling solvent floating zone technique [22]. Post-annealing procedures were carried out to obtain compositional uniform crystals. Structural characterizations were carried out by XRD. The chemical composition of the La–Bi2201 crystals was determined by induction-coupled plasma atomic emission spectroscopy (ICP-AES), and the actual composition is very close to their nominal composition. Resistivity, magnetic susceptibility and details of sample synthesis and characterization have been reported in the previous paper [22].

TEM investigations were performed on a Tecnai G2 F20 electron microscope operated at 200 kV. Cross-sectional TEM specimens were prepared by mechanical polishing and then ion milling at liquid nitrogen temperature, while plane-view specimens were cleaved using adhesive tape and then dissolved in chloroform solution and deposited on a carbon film suspended on a copper grid. *In situ* cooling TEM studies were carried out using the GATAN double tilt helium holder.

3. Results and discussions

3.1. Pristine Bi2201 phase

In order to give a direct comparison, we first carried out TEM measurements on the pristine phase Bi2201. For the pristine Bi2201, we mainly observed a monoclinic superstructure in $\boldsymbol{b}^*$-$\boldsymbol{c}^*$ plane, which is consistent with the previous reports [12, 13, 15]. The DPs and HRTEM images of both [001] and [100] zone axes are shown in figure 2. The main spots in DPs can be well indexed using the orthogonal unit cell with cell parameters a=5.362 Å, b=5.374 Å, c=24.622 Å. The modulation vector can be obtained directly from the DPs along [100] zone axis in figure 2 (c), $q_1 \approx 0.197b^* + 0.381c^*$. The component along $c^*$ is smaller than the previously reported values [13], which may be due to different Bi/Sr ratio of the samples. The $q_1$ determined from our study is very close to that in the Bi/Sr≈1 samples in the systematic study of **q**-dependence on Bi/Sr ratio [23]. The modulated structure schematically shown as the white lines can be clearly seen from the [100] and [001] zone axes HRTEM images. It is usually considered that the modulation is enhanced by the periodic distortions of the BiO bilayers [12, 24], which can also be seen from our HRTEM images. As shown in figure 2(d), the dark double contrast marked by arrows can be considered as BiO layers, and the contrast change shows the distortion of BiO atoms modulation resulting in the satellite spots shown in figure 2(a) and 2(c). This modulation may be affected by oxygen concentration. Recent theoretical investigations for $Bi_2Sr_2CaCu_2O_8$ indicated that spontaneous symmetry breaking alone without extra oxygen involved may also lead to the superstructure [25].

It is interesting to point out that, in the pristine Bi2201, there exists a minor second phase with a monoclinic structure. From DPs in figure 3 (a) and (c), the unit cell of the new phase can be determined as a≈5.4 Å, b≈5.7 Å, c≈25 Å and α≈119º. In the framework of this new unit cell, the supermodulation can be considered as a commensurate superstructure along $b^*$ axis, $q_1=0.25b^*$ and is illustrated by white lines in figure 3(d). This phase may result from a slight composition deviation from the ideal elemental ratio in Bi2201. It can be considered as a defect structure among the main phase in Bi2201. In fact, it is similar to the B-phase reported in [26] which was considered as $Bi_{17}Sr_{16}Cu_7O_y$.

3.2. La-doped Bi2201

We carried out systematic TEM studies on a serial of La-Bi2201 samples. Let us first focus on the $Bi_2Sr_{1.27}La_{0.73}CuO_6$ composition (x=0.73). From the [100] zone axis DP shown in figure 4(c), we found that the modulation in this La-Bi2201 (x=0.73) sample has changed into an orthogonal commensurate one from the monoclinic structure in the pristine Bi2201 phase. The modulation vector is determined to be $q_1=0.25b^*+c^*$. This orthogonal superstructure can also be seen from the HRTEM image, as marked schematically as white line square in figure 4(d). Along the *b* axis, 4-unit supercell of conventional unit cell can be constructed, while along the $c^*$ axis, a distortion makes the extinction 001 diffraction spot visible. The superstructure has a face centered feature and can still be considered to be originated from the distortions of the BiO bilayers. We also noted that the first satellite spot of $q_1$ is weak and may vanish in [001] zone axis DP in thin areas of sample.

From DPs along [001] zone axis, shown in figure 4(a), a new series of superstructure spots along the $b^*$ direction can be identified, which exhibits a new modulation $q_2=0.125b^*$. These satellite spots only appear around the extinction (2*n*+1 *k* 0) main spots while vanish around (2*n k* 0) main spots (*n, k* are integers). The origin of the modulation $q_2$ can be revealed from the HRTEM image along the [001] zone-axis, shown in figure 4(b). A pair of four basic cells corresponding to the $q_1$ modulation forms an anti-phase structure that is eight units along *b* of basic structure (~4.3nm). Such structure can be schematically shown as the model structure in the inset of figure 4(b). Black and white circles (and squares) illustrate the phase change of π that results in the contrast reversal, while the rows of squares demonstrate the distortions corresponding to the $q_1$ modulation. From such a model, we know that the $q_2$ modulation is directly evolved from the orthogonal modulation $q_1$ through an addition of the anti-phase structure. The superstructure $q_2$ was in fact observed in La-Bi2201 (x=0.3) sample before [18], but it was not explicitly pointed out. We also noted that any extra spots corresponding to this $q_2$ superstructure cannot be seen from the [100] and [010] zone axes DPs. Therefore, such a superstructure may come from additional displacements along *a* direction, that is, the $q_2$ modulation is a transverse wave directing along *b* axis in the *ab* plane.

We have carried out a series of electron diffraction studies on the superstructure evolution in La-Bi2201 with La concentration varying from x=0.25 to 1.10, corresponding to a doping level change from overdoped to underdoped and even to non-superconducting samples, as shown in figure 5. From the DPs along the [001] zone axis, we can see that all the La-Bi2201 samples have such satellite spots corresponding to the anti-phase structure and the modulus of the vector $q_2$ exactly equals to half of corresponding component of $q_1$ along $b^*$ direction in the entire doping range from x=0.25 to 1.10 samples. Table 1 summarizes the evolution of the $q_1$ and $q_2$ values with the La concentration x. The component of vector $q_1$ along $b^*$ shows a slight increase when x changes from 0.25 to 0.84. However, when the nominal La-concentration x=1.10, the component of vector $q_1$ along $b^*$ dramatically increased to 1/3.5 (~0.284) $b^*$ which is close to previous reported value 1/3.4 $b^*$ in La-Bi2201 (x=1.0) sample [12]. It is worth mentioning that the pristine Bi2201 only shows $q_1$ superstructure modulation without any indication of $q_2$. However, the new $q_2$ modulation appears in all the La-Bi2201 samples, regardless of the La content. Moreover, the modulation $q_1$ and $q_2$ are directly coupled. As it is usually considered, the $q_1$ superstructure in Bi2201 originates from the distortions in the BiO double layers. Substituting $La^{3+}$ for $Sr^{2+}$ will introduce extra electrons and thus change the charge balance of the BiO layers; this may change the modulation center and result in the unusual orthogonal commensurate superstructure $q_1$. The appearance of the $q_2$ modulation is clearly related with the La-substitution. Because the atomic radius of $La^{3+}$ is smaller than $Sr^{2+}$, $La^{3+}$ substitution for $Sr^{2+}$ may cause additional atom displacement that gives rise to the anti-phase stripe structure that we have observed.

In order to further clarify the characteristics of the newly found $q_2$ superstructure, we have carried out *in situ* temperature-dependent TEM studies on the La-Bi2201(x=0.73) sample with a helium cold TEM holder. We have measured DPs of [001] zone axis from room temperature down to 30 K and found that both $q_1$ and $q_2$ show little change with temperature.

The $q_2$ modulation observed in our TEM studies clearly indicates that it is associated with the bulk property. Our TEM $q_2$ is different from the LEED $q_2$ reported before [20] in a number of aspects. First, they show different composition dependence. In the LEED experiments, the second supermodulation $q_2$ can be observed in underdoped La-Bi2201 (x=0.75, 0.8) samples but not in optimal doped samples (x=0.5). This is different from our TEM superstructure $q_2$ that exists in the entire composition range from x=0.25 to x=1.1. Second, they show different temperature dependence. It was reported that the LEED $q_2$ changes dramatically with temperature, from about $q_1/3$ above 130K to $q_1/2$ at 6K [21]. But our TEM $q_2$ shows little change with temperature from room temperature down to 30K. However, we note that the temperature dependence of our TEM $q_2$ is similar to the $q_2$ observed by XRD and resonant x-ray scattering (REXS) experiments [21]. It remains to be investigated whether the surface $q_2$ from LEED and the bulk $q_2$ from our TEM and previous XRD have the same origin, and why they show different doping and temperature dependences.

Our present TEM study of the new $q_2$ modulation will also provide insight on the origin of the Fermi pocket recently observed in underdoped La-Bi2201 [1]. It is highly unlikely that the Fermi pocket observed in ARPES [1] can be explained by the new superstructure modulation $q_2$ observed in our TEM. In the ARPES measurements, the Fermi pocket was observed only in the underdoped region, but not in the optimally-doped and overdoped samples [1]. This is apparently inconsistent with the TEM $q_2$ that is present in the entire composition range of x=0.25~1.1 that covers the doping level from overdoped to optimally-doped, to heavily underdoped samples. We also note that the modulation of the anti-phase stripe superstructure we observed is a secondary effect on the commonly known modulation $q_1$. This modulation is very weak in nature and whether it may give an observable effect on the electronic structure in the ARPES measurements is a question that needs further investigations.

4. Conclusions

The microstructures of the La-Bi2201 samples covering a wide range of composition, x=0~1.1, have been systematically studied by electron diffraction and HRTEM images. We confirmed the monoclinic superstructure and determined the modulation vector in pristine phase Bi2201. In the La-Bi2201 (x=0.73) sample, the monoclinic superstructure has changed into an orthogonal face centered commensurate structure and the modulation vector is $q_1=0.25b^*+c^*$. The modulation vector along $b^*$ direction $q_{1b}$ increases slightly from $0.24b^*$ to $0.246b^*$ when the La-concentration x increases from 0.25 to 0.84. However, when x=1.10, the $q_{1b}$ increases dramatically to about $0.286b^*$. A new anti-phase superstructure with modulation vector $q_2=q_{1b}/2$ emerged in our entire composition range of x=0.25~1.10. The new $q_2$ superstructure is closely related to the $q_1$ superstructure commonly observed; both of them show little change with temperature from room temperature to 30 K. In terms of doping and temperature dependence, this anti-phase $q_2$ superstructure we observed in TEM is different from the incommensurate $q_2$ superstructure recently observed in LEED and ARPES experiments. Whether this anti-phase $q_2$ is the origin of the Fermi pocket formation observed in the underdoped La-Bi2201 in the ARPES measurements needs further investigations.

Acknowledgements

This work was supported by the National Science Foundation of China under Contracts No. 10874227, No. 90922001, the Knowledge Innovation Project of the Chinese Academy of Sciences, and the 973 projects of the Ministry of Science and Technology of China. XJZ thanks the funding support from NSFC (Grant No. 11190022) and the MOST of China (Program No: 2011CB921703 and 2011CB605903).

Table 1. Evolution of **q₁** along the $b^*$ direction ($q_{1b}$) and **q₂** with the La concentration x in La-Bi2201. The error bar of $q_{1b}/b^*$ is smaller than 0.01.

Figure captions

Figure 1. Basic crystal structure of $Bi_2Sr_{2-x}La_xCuO_6$. Unit cell parameters are a=5.362Å, b=5.374 Å, c=24.622 Å. The $CuO_2$ plane is also schematically shown in the figure.

Figure 2. Electron DPs and HRTEM images of Bi2201 pristine phase. (a) Selected area DP along the [001] zone-axis; (b) [001] zone-axis HRTEM image; (c) selected area DP along the [100] zone-axis; (d) [100] zone-axis HRTEM image. **q₁** illustrates the supermodulation, and the arrows in (d) mark the BiO bilayers. The monoclinic supercell is also indicated by white lines in (d).

Figure 3. Electron DPs and HRTEM images of a defect structure in the Bi2201 pristine phase. (a) Selected area DP along the [001] zone-axis; (b) [001] zone-axis HRTEM image; (c) selected area DP along the [100] zone-axis; (d) [100] zone-axis HRTEM image.

Figure 4. Electron DPs and HRTEM images of $Bi_2Sr_{1.27}La_{0.73}CuO_6$ (La-Bi2201, x=0.73). (a) Selected area DP along the [001] zone-axis. **q₂** indicates an anti-phase modulation vector; (b) [001] zone-axis HRTEM image. The upper-right inset shows the structure model illustrating the anti-phase superstructure; (c) selected area DP along the [100] zone-axis; (d) [100] zone-axis HRTEM image. The white line squares are schematically shown as the supercell.

Figure 5. Electron DPs along the [001] zone-axis as $0<x\leqslant1.10$ in $Bi_2Sr_{2-x}La_xCuO_6$, (a) x=0.25, (b) x=0.40, (c) x=0.60, (d) x=0.73, (e) x=0.84, (f) x=1.10. The double white and black arrows denote the satellite spots corresponding to the anti-phase superstructure 2**q₁** and 2**q₂** respectively.

Table 1.

| | La concentration x in $Bi_2Sr_{2-x}La_xCuO_6$ | | | | | | |
|---|---|---|---|---|---|---|---|
| | 0 | 0.25 | 0.40 | 0.60 | 0.73 | 0.84 | 1.10 |
| $q_{1b}/b^*$ | 0.197 | 0.240 | 0.235 | 0.241 | 0.249 | 0.246 | 0.286 |
| $q_2/b^*$ | --- | 0.120 | 0.113 | 0.120 | 0.125 | 0.123 | 0.143 |

Figure 1.

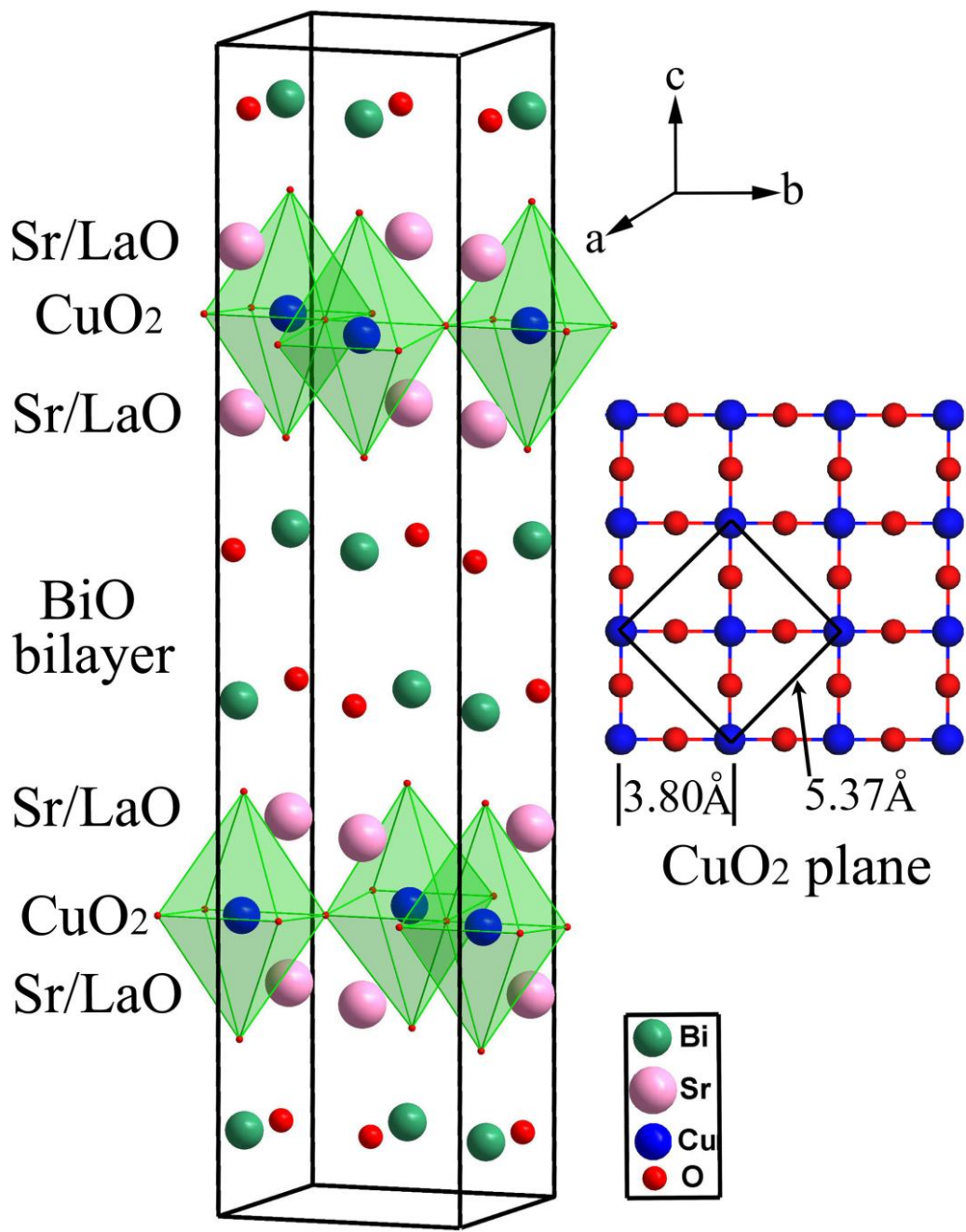

Figure 2.

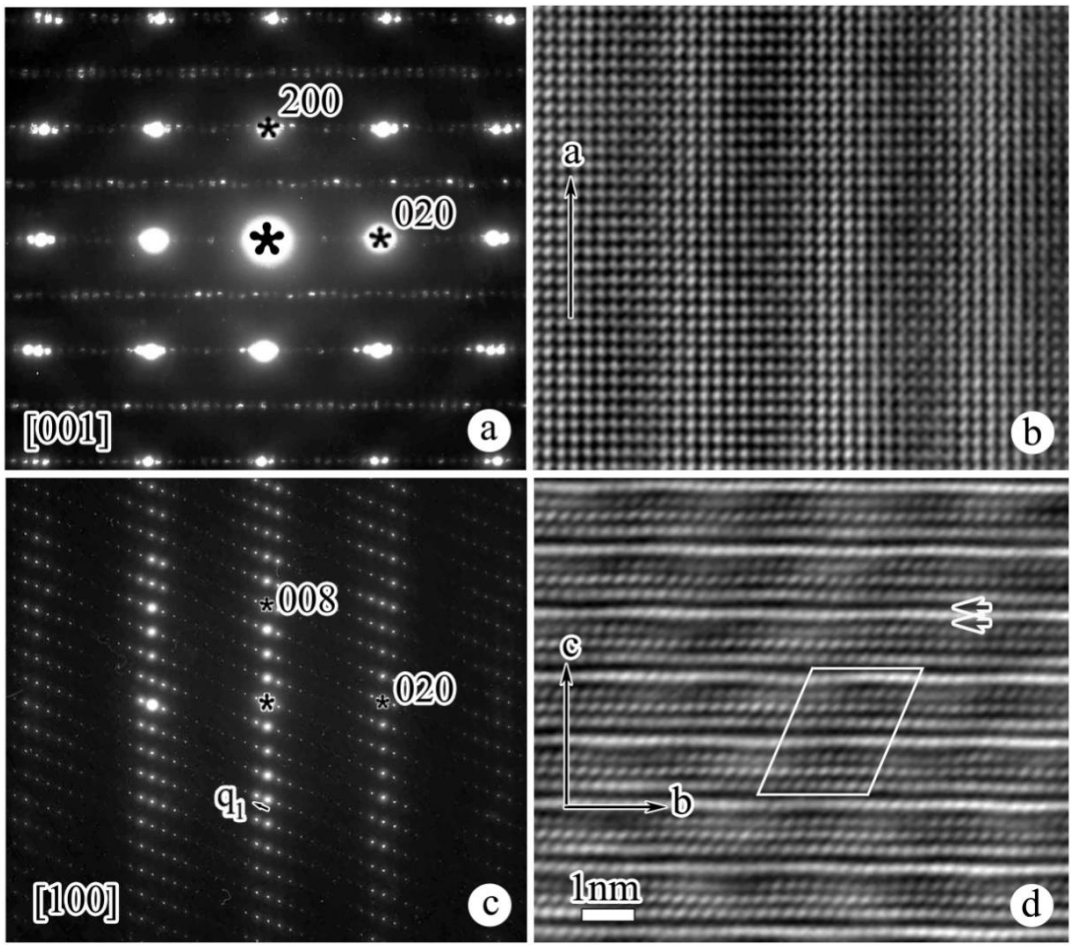

Figure 3.

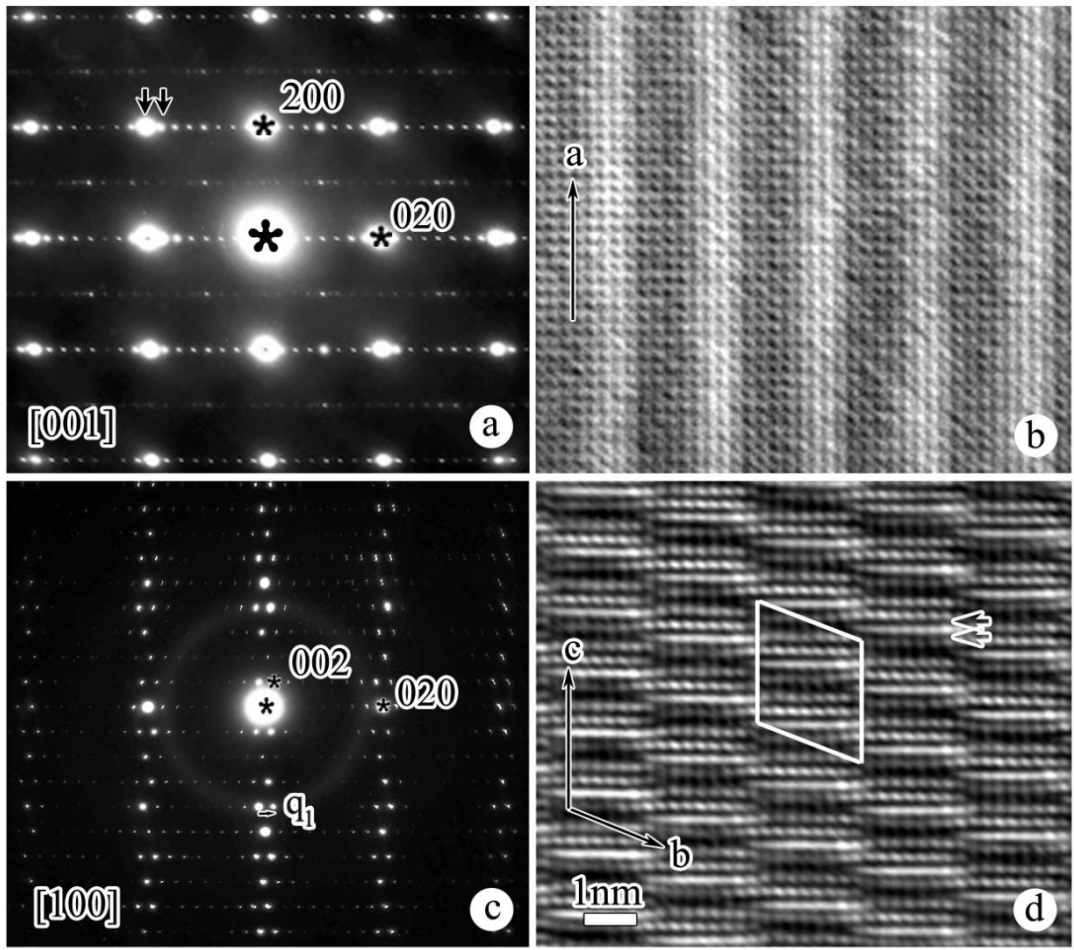

Figure 4.

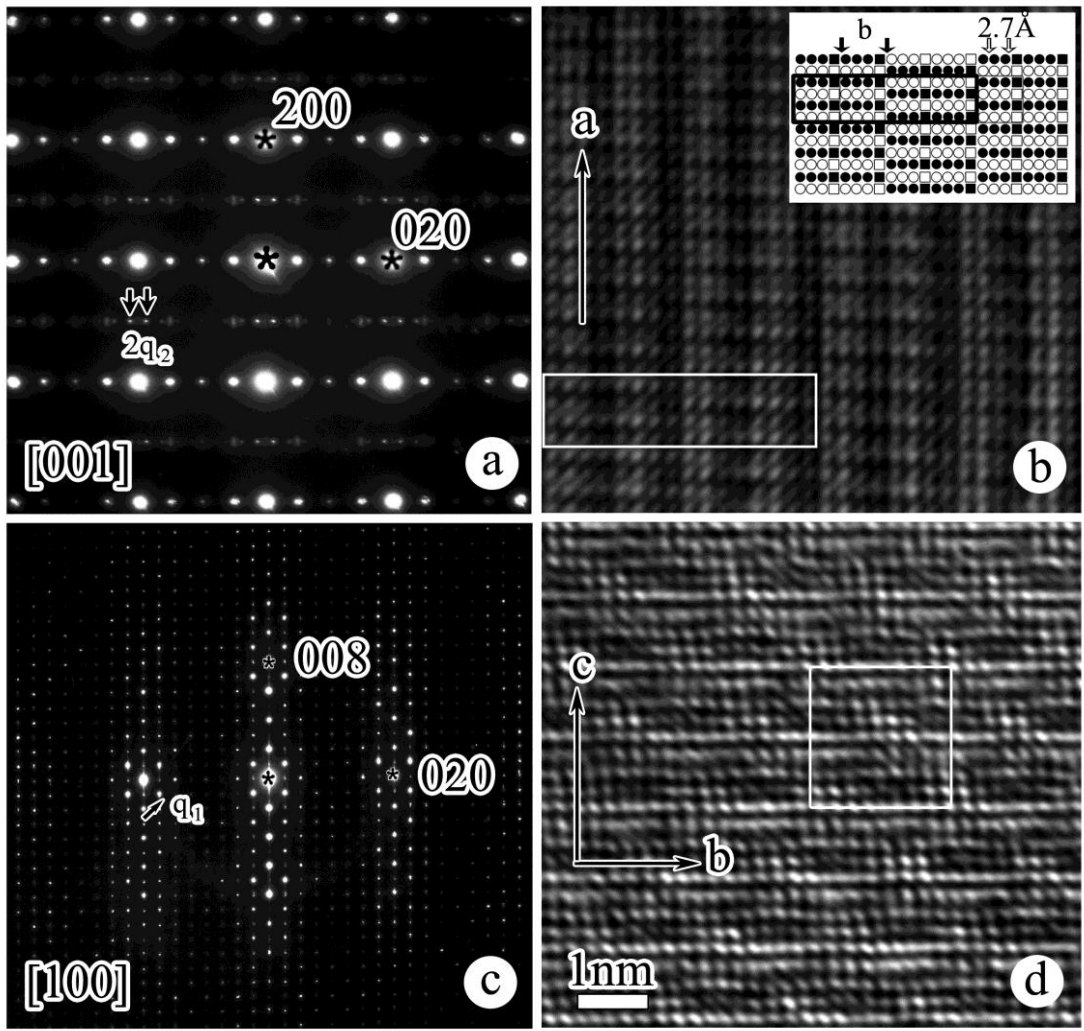

Figure 5.

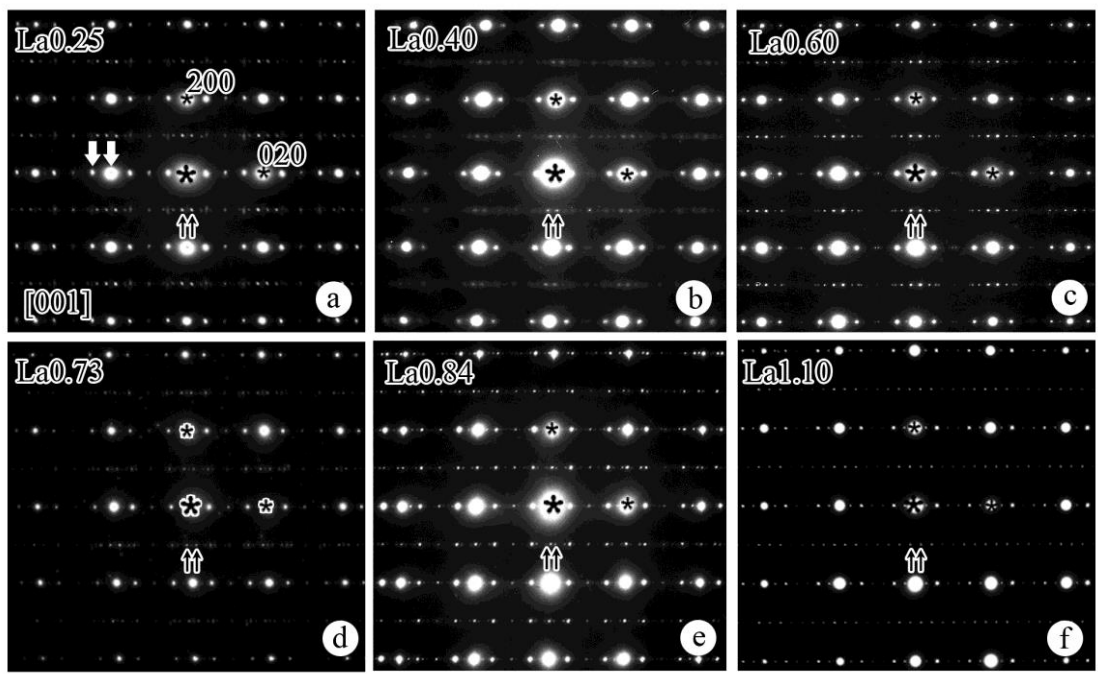